\documentclass[final,5p,number,sort&compress]{elsarticle}
\usepackage{lineno}
\usepackage{float}
\usepackage{hyperref}
\usepackage{amsmath}
\usepackage{amssymb}

\usepackage{amsthm}
\usepackage{enumerate}
\usepackage{cleveref}
\usepackage{physics}
\usepackage{mhchem}
\usepackage{braket}

\journal{Nuc.\ Instrum.\ Meth.}

\begin{document}

\hyphenation{CACTUS}
\hyphenation{outpopulation y}

\begin{frontmatter}

\title{RAINIER: A Simulation Tool for Distributions of Excited Nuclear States and Cascade Fluctuations}

\author[berk]{\corref{cor1}L. E. Kirsch}
\ead{kirsch2@berkeley.edu}
\author[berk,lbnl]{L. A. Bernstein}
\address[berk]{Nuclear Engineering, UC Berkeley, CA 94720, USA.}
\address[lbnl]{Lawrence Berkeley National Laboratory, Berkeley, CA 94720, USA}

\cortext[cor1]{Corresponding Author}

\begin{abstract}
A new code has been developed named \texttt{RAINIER} that simulates the $\gamma$-ray decay of discrete and quasi-continuum nuclear levels for a user-specified range of energy, angular momentum, and parity including a realistic treatment of level spacing and transition width fluctuations.
A similar program, \texttt{DICEBOX}, uses the Monte Carlo method to simulate level and width fluctuations but is restricted to $\gamma$-ray decay from no more than two initial states such as de-excitation following thermal neutron capture.
On the other hand, modern reaction codes such as \texttt{TALYS} and \texttt{EMPIRE} populate a wide range of states in the residual nucleus prior to $\gamma$-ray decay, but do not go beyond the use of deterministic functions and therefore neglect cascade fluctuations.
This combination of capabilities allows \texttt{RAINIER} to be used to determine quasi-continuum properties through comparison with experimental data.
Several examples are given that demonstrate how cascade fluctuations influence experimental high-resolution $\gamma$-ray spectra from reactions that populate a wide range of initial states.
\end{abstract}

\begin{keyword}
\texttt{simulation \sep Monte-Carlo \sep gamma cascade \sep reaction \sep initial distribution}
\end{keyword}

\end{frontmatter}

\section{\label{sec:intro}Introduction}

The modeling of nuclear reaction rates for a broad range of applications including astrophysical nucleosynthesis, counter-proliferation, and stewardship science requires accurate knowledge of the properties of highly-excited nuclear states near the particle separation energy.
There is a large and growing set of experimental data using the Oslo \cite{GUTTORMSEN1987518}, $\beta$-Oslo \cite{PhysRevLett.113.232502}, and Direct Reaction Two Step Cascade (DRTSC) \cite{PhysRevLett.108.162503} methods which can be used to inform models of these excited states.
Furthermore, the advent of new high-resolution event tracking $\gamma$-ray spectrometers such as GRETINA \cite{Paschalis201344} and AGATA \cite{AKKOYUN201226} offer the possibility of providing direct insight into the transition widths of these states through the use of lifetime measurements via the Doppler Shift Attenuation Method (DSAM) \cite{Alexander1978}. 
However, the interpretation of data from all these experiments requires the use of a $\gamma$-ray cascade model that simulates an initial state population covering a wide range of $EJ\Pi$ while also incorporating a realistic treatment of level spacing and transition width fluctuations.

The statistical nuclear decay code \texttt{DICEBOX} \cite{Bečvář1998434} uses a Monte Carlo approach to create and decay levels, naturally incorporating level spacing and transition width fluctuations. 
However, the authors of \texttt{DICEBOX} developed the code to describe Two Step Cascades (TSC) following thermal neutron capture \cite{PhysRevC.46.1276} which only populates two initial states separated by $1\hbar$.
Therefore, \texttt{DICEBOX} is not appropriate for modeling the decay of a nucleus populated in $\beta$-decay, transfer reactions, or high-energy compound reactions.
In contrast, the nuclear reaction codes \texttt{TALYS} \cite{TALYS04} and \texttt{EMPIRE} \cite{HERMAN20072655} sample a wide range of $EJ\Pi$ but deterministically model the $\gamma$-ray cascades of residual nuclei with nothing more than smooth level density and transition width functions above an energy threshold.
In reality, fluctuations in the Nuclear Level Density (NLD) and the Gamma Strength Function (GSF) play an important role in low-lying discrete state populations.

This paper describes a new \texttt{C++} program, the \textbf{R}andomizer of \textbf{A}ssorted \textbf{I}nitial \textbf{N}uclear \textbf{I}ntensities and \textbf{E}missions of \textbf{R}adiation (\texttt{RAINIER}), that incorporates a Monte Carlo construction of nuclear level structure with the ability to populate a set of states spanning a wide range of $EJ\Pi$, thereby enabling the interpretation of discrete state population data to inform nuclear structure models in the quasi-continuum.
\texttt{RAINIER} only allows for decay via $\gamma$-ray emission or internal conversion and is therefore appropriate only for bound states.
\texttt{RAINIER} opens the possibilities of testing experimental techniques such as the Oslo Method \cite{GUTTORMSEN1987518}, generating feeding time distributions for studies of quasi-continuum lifetimes, and using observed discrete state populations to determine a nucleus's underlying angular momentum distribution.

\section{\label{sec:method}Method}

\texttt{RAINIER}'s intended use is for modeling $\gamma$-ray cascades only (e.g., following emission of the last massive particle).
\texttt{RAINIER} takes the following steps to simulate the complete, high-resolution $\gamma$-ray spectra from the residual nucleus:
\begin{itemize}
\item Build the low-energy portion of the level scheme from available information in structure databases
\item Use NLD models to construct the upper portion of the level scheme. This set of artificially generated discrete levels is known as a nuclear ``realization''
\item Populate a user-specified distribution of initial levels
\item Depopulate levels using GSF models
\item Compute and histogram quantities such as emitted $\gamma$-ray energies, level populations, and decay times
\end{itemize}
These steps are described in greater detail in the following sections.

\begin{figure}
\centerline{\includegraphics[width=0.75\linewidth]{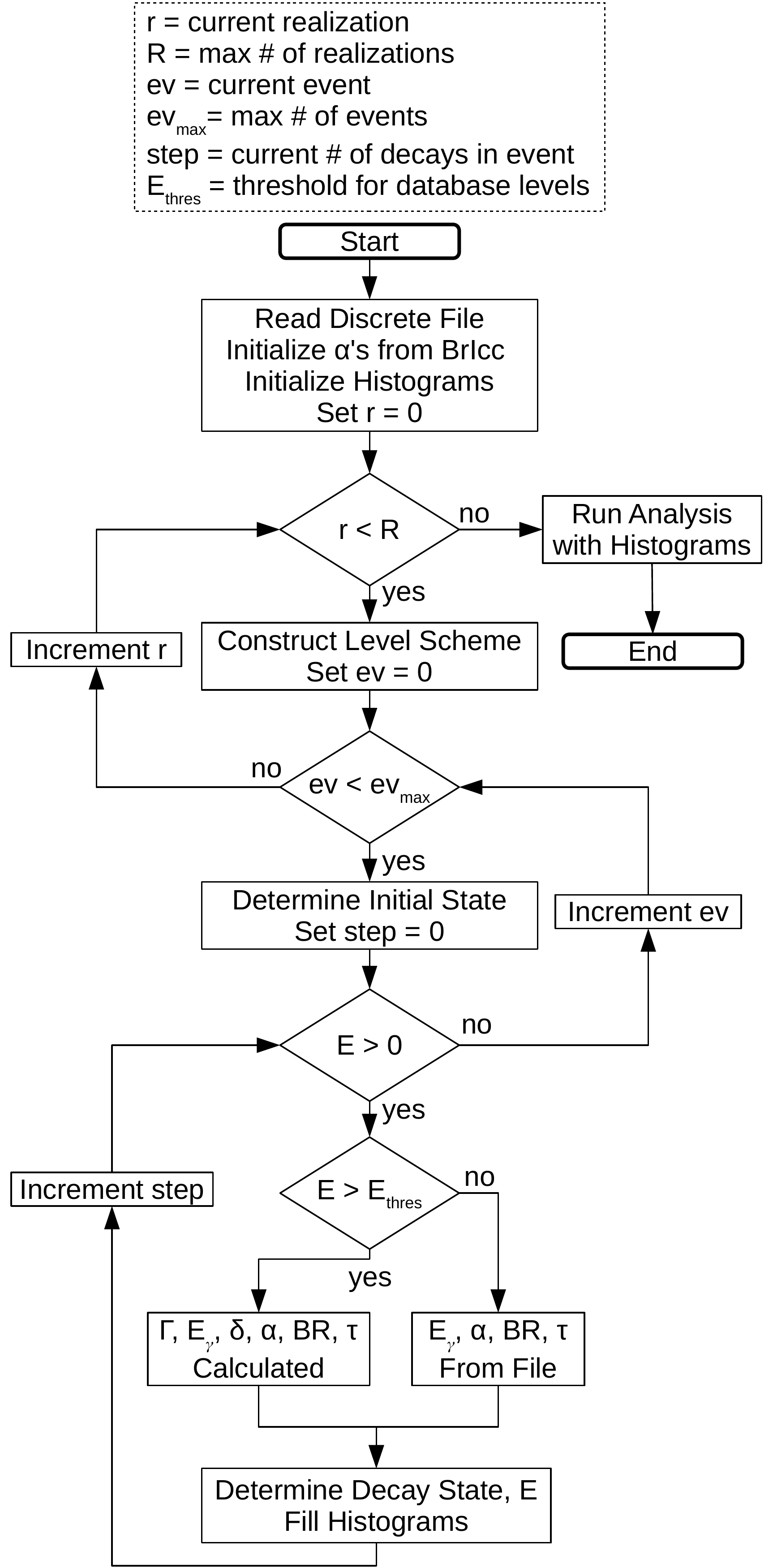}}
\caption{
Program flow in \texttt{RAINIER}.
Physical variables $\Gamma$, $\delta$, $\alpha$, $BR$, and $\tau$ described in text. }
\label{fig:flow}
\end{figure}

Figure \ref{fig:flow} shows the execution order of \texttt{RAINIER}.
To achieve low statistical uncertainty, users set the maximum number of events, $ev_\textrm{max}$, large enough to obtain many instances of a desired observable.
Users also set the maximum number of nuclear realizations, $R$, to track the influence of level spacing and width fluctuations on that observable.

\subsection{\label{ssec:build}Constructing the Level Scheme}

The Reference Input Parameter Library (RIPL-3) \cite{CAPOTE20093107} supplies level information below a user-defined energy threshold, $E_{thres}$, below which \texttt{RAINIER} does not generate levels.
For each level below $E_{thres}$, \texttt{RAINIER} reads $EJ\Pi$, lifetime, $\tau$, and all $\gamma$-ray decay exit channels with corresponding branching ratio, $BR$, and total internal conversion coefficient, $\alpha$.

The region above $E_{thres}$, referred to as the \textit{constructed level scheme}, depends on NLD models.
The default total NLD model is the Back Shifted Fermi Gas (BSFG) model \cite{doi:10.1139/p56-090,DILG1973269}:
\begin{equation}
\rho_{\textrm{BSFG}}^{tot}(E) = \frac{1}{12 \sqrt{2} \sigma} 
  \frac{ \textrm{exp}\left(2 \sqrt{a U}\right) } {a^{1/4} U^{5/4} },
\label{eq:BSFG}
\end{equation}
where $U = E - E_1$ is the effective excitation energy, $E_1$ is the energy backshift, $a$ is the level density parameter related to orbital energy spacings, and $\sigma$ is the spin cutoff parameter.
A constant temperature model of total NLD \cite{ERICSON1959481,GilCam} is also available: 
\begin{equation}
\rho_T^{tot}(E) = \frac{1}{T_0} \textrm{exp}\left( \frac{E - E_0}{T_0} \right),
\label{eq:CTM}
\end{equation}
where $E_0$ is backshift and $T_0$ is temperature.
Von Egidy and Bucurescu \cite{PhysRevC.72.044311,PhysRevC.73.049901,PhysRevC.80.054310} provide tables of $E_1$, $a$, $E_0$, and $T_0$ from empirical fits of complete level schemes at low excitation energies in combination with $s$-wave neutron resonance spacings at the neutron binding energies.
Von Egidy and Bucurescu also provide global equations involving only quantities available from mass tables for extrapolation to nuclei where there is insufficient data.
\texttt{RAINIER} also offers an energy dependent form of $a$:
\begin{equation}
a(E) = \tilde{a} \bigg[1 + W \frac{1 - \textrm{exp}(-d \cdot U)}{U} \bigg],
\end{equation}
where $\tilde{a}$ is the asymptotic value of $a$ devoid of shell effects, $d$ is the damping parameter, and $W$ is the shell correction energy.
Typically the energy dependence of $a$ is omitted since it does not play a major role below 10 MeV. 

A Fermi gas model determines the underlying $J$ distribution \cite{EricsonStatModLvlDen}:
\begin{equation}
R_F(E, J) = \frac{2 J + 1}{2 \sigma^2} \textrm{exp}\left[- \frac{(J + 1/2)^2}{2 \sigma^2} \right].
\label{eq:spDist}
\end{equation}
Theoretical versions of the energy dependent spin cutoff parameter are available including a low-energy model \cite{theoSpinCut}
\begin{equation}
\sigma^2 = 0.0146 A^{5/3} \frac{1 + \sqrt{1 + 4 a U} }{2 a},
\label{eq:spinLowE}
\end{equation}
a single-particle states model \cite{PhysRevC.75.044308}
\begin{equation}
\sigma^2 = 0.1461 \sqrt{a U} A^{2/3},
\label{eq:spinSing}
\end{equation}
and a rigid sphere model \cite{PhysRevC.10.2373}
\begin{equation}
\sigma^2 = 0.0145 \sqrt{U/a} A^{5/3},
\label{eq:spinRig}
\end{equation}
where $A$ is the atomic mass number.
The empirical version of spin cutoff from von Egidy and Bucurescu \cite{PhysRevC.80.054310} is also available:
\begin{equation}
\sigma^2 = 0.391 A^{0.675} (E - Pa')^{0.312},
\label{eq:spinEmpirical}
\end{equation}
where $Pa'$ is the deuteron pairing energy calculated from mass tables \cite{AUDI2003337}.

The default parity distribution is parity equipartition:
\begin{equation}
\pi(E, J, \Pi) = 1/2.
\label{eq:par}
\end{equation}
An energy dependent parity distribution \cite{PhysRevC.67.015803} is also available:
\begin{equation} 
\pi(E, J, \Pi) = \frac{1}{2} \left(1 \pm \frac{1}{1 + \textrm{exp}[C (E - D)] } \right),
\end{equation}
where $C$ and $D$ are free parameters and the $\pm$ symbol depends on the sign of $\Pi$ and whether $A$ is even or odd.

Together, the three components of NLD are
\begin{equation}
\rho(E, J, \Pi) = \rho_T^{tot}(E) \hspace{1mm} R_F(E, J) \hspace{1mm} \pi(E, J, \Pi),
\label{eq:lvlEJP}
\end{equation}
which represents the number of nuclear levels near $E$ for a given $J$ and $\Pi$.

The even-odd $J$ staggering of even-even nuclei, likely related to the pairing interaction, is omitted since it is predicted to disappear at large excitation energies \cite{PhysRevLett.99.162504,PhysRevC.78.051301}.

Users define a maximum energy $E_{max}$, an energy bin spacing $\delta E$, and a maximum angular momentum bin $J_{b,max}$ of the constructed level scheme.
Bins of $J$ are separated by one unit of angular momentum and can take on integer or half-integer values.
These restrictions fully bound and pixelate the constructed level scheme into $EJ\Pi$ bins.
\texttt{RAINIER} gives each level within a bin its own unique identity independent of bin number or content.
The level generation algorithm is the only program routine that acknowledges the existence of energy bins.

Following pixelization of the constructed level scheme, \texttt{RAINIER} randomly generates level contents for each bin according to a Poisson distribution:
\begin{equation}
P(n) = e^{-\lambda} \lambda^n / n!
\label{eq:poisson}
\end{equation}
where $n$ is an \textit{integer} number of levels in the $EJ\Pi$ bin and 
\begin{equation}
\lambda = \delta E \cdot \rho(E, J, \Pi)
\label{eq:lvlFluct}
\end{equation}
is the \textit{real-valued} expected number of levels with $E$ as the centroid value of the energy bin.
\texttt{RAINIER} provides a second option for bin content generation from Random Matrix Theory \cite{tagkey2004ifc1} where the distance between levels, $Q$, follows a Wigner distribution:
\begin{equation}
P(q) = \frac{1}{2} \pi q e^{-\pi q^2 /4},
\label{eq:lvlSpac}
\end{equation}
where $q = Q \cdot \rho(E, J, \Pi)$ is the reduced level spacing.
\texttt{RAINIER} generates all levels once at the beginning of each realization.

Since constructed level scheme fluctuations are a concern, \texttt{RAINIER} can generate several different realizations of the level scheme and map variations in output observables.
Larger values of $E_{max}$, a larger number of realizations, $R$, and a larger number of events, $ev_\textrm{max}$, increase program execution time.

\subsection{\label{ssec:ipop}Initial Level Population}

There are many different ways to experimentally populate levels in a given nucleus.
For example, thermal neutron capture predominantly populates a single state near the neutron separation energy with $J$ equal to the ground state angular momentum of the target nucleus $\pm$ $ \hbar / 2$.
In contrast, $\beta$-decay populates a small selection of states primarily with difference in $J$ less than two units of angular momentum from the parent nucleus.
Inelastic scattering reactions such as (p,p') bring in a range of angular momentum depending on the angle of the emitted particle.
Heavy ion fusion reactions such as ($^{48}$Ca,xn) supply a lot of angular momentum and populate states almost exclusively along the yrast and yrare bands. 

\begin{figure}[h]
\centerline{\includegraphics[width=0.48\linewidth]{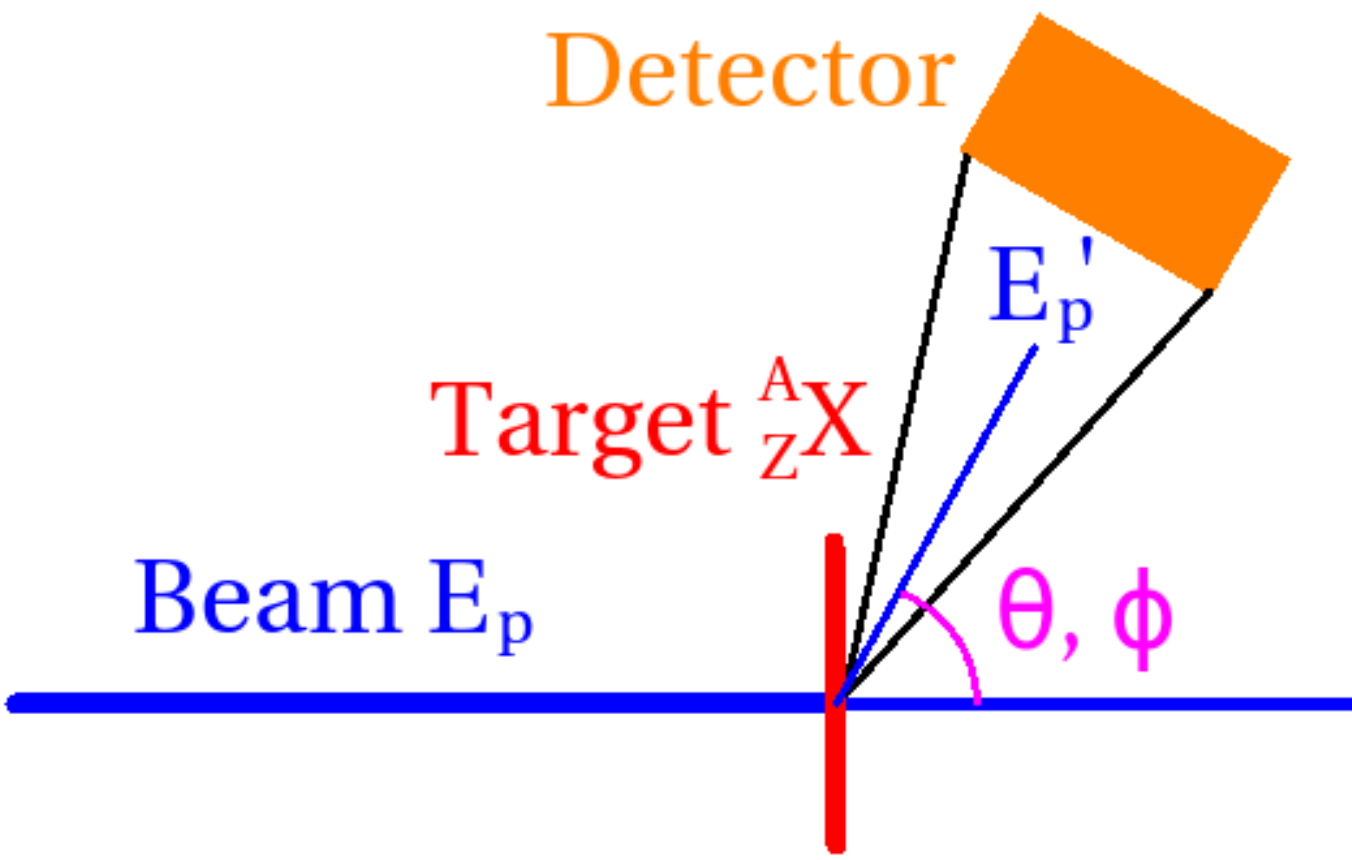}}
\caption{Observation constrained initial nuclear states.}
\label{fig:ExSpread}
\end{figure}

Initial states are typically experimentally constrainable as demonstrated in Figure \ref{fig:ExSpread}.
Adequate proton energy resolution in (p,p') limits initial excitation energy, $E_{I}$.
Proton emission angle with respect to the beam axis, $\theta$, may limit initial angular momentum, $J_{I}$.
Azimuthal scattering angle, $\phi$, of an incident \textit{polarized} beam may limit initial parity, $\Pi_I$.
In data analysis, one can select $\gamma$-ray decay events from a specific set of experimentally constrained initial excitations and compare to \texttt{RAINIER} decay simulations of similar initial states.

To address the different types of experimental constraints, \texttt{RAINIER} has the following built-in initial state population modes:
\begin{enumerate}
  \item \textit{single state}; akin to (n,$\gamma$)
  \item \textit{selection of states} of varying probabilities; akin to $\beta$-decay
  \item \textit{spread of states}; akin to ejectile energy constrained inelastic scattering
  \item \textit{full reaction} from $EJ\Pi$ histogram; akin to heavy ion fusion
\end{enumerate}
With these operation modes the user can also simulate photoabsorption, $\alpha$-decay, neutron pickup ($^3$He,$\alpha$) reactions, spallation, and many more experimental scenarios.
The histogram used in the \textit{full reaction} operation mode is a typical output of reaction codes like \texttt{TALYS} and \texttt{EMPIRE} for which \texttt{RAINIER} can effectively perform the final stage processing.

\subsection{\label{ssec:width}Transition Widths}

\texttt{RAINIER} applies the extreme statistical model postulated by Bohr \cite{Bohr:1936zz} that assumes the decay of a nuclear level is independent of the way in which it is formed. 
For example, the transition width of level $A$ to $B$ is unchanged if inelastic scattering directly populates $A$ or decay from high-lying state $C$ indirectly populates $A$.

Average transition width is related to the GSF, by \cite{Bartholomew1973}
\begin{equation}
\bar{\Gamma}^{XL}(E,E_\gamma) = \frac{f_{XL}(E_\gamma) E_\gamma^{2 L + 1}} 
{ \rho(E,J,\Pi)},
\label{eq:avgWidth}
\end{equation}
where $X$ is the transition electromagnetic character, $L$ is the transition multipolarity, $E_\gamma$ is the energy of the emitted $\gamma$-ray, and $f_{XL}(E_\gamma)$ is the GSF of transition type $XL$, assumed independent of $J$ and $\Pi$ according to the Brink hypothesis \cite{brink1955phd}.

In \texttt{RAINIER}, the default M1 GSF is a standard Lorentzian \cite{BRINK1957215,PhysRev.126.671}:
\begin{equation}
f_{XL}(E_\gamma) = K_{XL}
  \frac{S_{XL} E_\gamma G_{XL}^2}
  { ( E_\gamma^2 - E_{XL}^2 )^2 + E_\gamma^2 G_{XL}^2}
\label{eq:GSFStd}
\end{equation}
where $S_{XL}, E_{XL}$, and $G_{XL}$ are the magnitude, centroid energy, and width of the giant resonance, respectively and
\begin{equation}
K_{XL} = \frac{1}{(2 L + 1) \pi^2 \hbar^2 c^2}.
\end{equation}
The M1 giant resonance parameters are given by 
\begin{align}
E_{M1} &= 41 \cdot A^{1/3} \hspace{2mm} \textrm{MeV} \\
G_{M1} &= 4 \hspace{2mm} \textrm{MeV} \\
f_{M1}( 7 \hspace{1mm} \textrm{MeV} ) &= 1.58 \times 10^{-9} \cdot A^{0.47} \hspace{2mm} \textrm{mb},
\label{eq:M2Param}
\end{align}
where Equation \ref{eq:GSFStd} is applied at 7 MeV to obtain the value $S_{M1}$.

The default E1 GSF is a Generalized Lorentzian (GLO) of the form of Kopecky and Uhl \cite{PhysRevC.41.1941}:
\begin{equation}
\begin{split}
f&_{XL}(E_\gamma,T) =  K_{XL} S_{XL} G_{XL} \\
& \times \Bigg[ F_\ell \frac{G_{XL} 4 \pi^2 T^2}{E_{XL}^5} 
+ \frac{E_\gamma \widetilde{G}_{XL}(E_\gamma,T) }{(E_\gamma^2- E_{XL}^2)^2 + E_\gamma^2 \widetilde{G}_{XL}(E_\gamma,T)^2} \Bigg] ,
\end{split}
\label{eq:GSFGen}
\end{equation}
where $F_\ell = 0.7$ is derived from the Fermi theory of liquids taking into account collisions between quasiparticles. 
The energy-dependent damping width is
\begin{equation}
\widetilde{G}_{XL}(E_\gamma,T) = G_{XL} \frac{E_\gamma^2 + 4\pi^2 T^2} {E_{XL}^2},
\label{eq:dampWid}
\end{equation}
and nuclear temperature is
\begin{equation}
T = \sqrt{\frac{E - E_1 - E_\gamma}{a}}.
\end{equation}
Other models for the E1 GSF are available including the KMF model \cite{KMF}:
\begin{equation}
f_{XL}(E_\gamma,T) = K_{XL} S_{XL} G_{XL} \cdot
F_\ell \frac{E_{XL} \widetilde{G}_{XL}(E_\gamma,T) }{(E_\gamma^2 - E_{XL}^2)^2}
\label{eq:GSFKMF}
\end{equation}
the model by Kopecky and Chrien \cite{KOPECKY1987285}:
\begin{equation}
\begin{split}
f_{XL}(E_\gamma,&T) =  K_{XL} S_{XL} G_{XL} \\
&\times \frac{E_\gamma \widetilde{G}_{XL}(E_\gamma,T)}{(E_\gamma^2 - E_{XL}^2)^2 + E_\gamma^2 \widetilde{G}_{XL}(E_\gamma,T)^2}
\end{split}
\end{equation}
and the Enhanced Generalized Lorentzian (EGLO) \cite{CAPOTE20093107} with the following modification to the damping width of Equation \ref{eq:dampWid} for $A > 148$:
\begin{align}
\widetilde{G}_{EGLO} &= \widetilde{G}_{XL} \cdot \chi(E_\gamma) \\
\chi(E_\gamma) &= k + (1-k) (E_\gamma - \epsilon_0) / (E_{E1} - \epsilon_0) \\
k &= 1 + 0.09 (A - 148)^2 \textrm{exp}[-0.18 (A -148) ] \\
\epsilon_0 &= 4.5 \hspace{2mm} \textrm{MeV}.
\label{eq:EGLO}
\end{align}

The E2 GSF can either be a single particle model of constant strength or a standard Lorentzian parameterized by the isoscalar mode of Prestwich \textit{et.\ al.\ }\cite{Prestwich1984,0034-4885-44-7-002}:
\begin{align}
E_{E2} &= 63 \cdot A^{1/3} \hspace{2mm} \textrm{MeV} \\
G_{E2} &= 6.11 - 0.012 \cdot A \hspace{2mm} \textrm{MeV} \\
S_{E2} &= 1.5 \times 10^{-4} Z^2 E_{E2} / ( A^{1/3} G_{E2} ) \hspace{2mm} \textrm{mb}.
\label{eq:E2Param}
\end{align}

For nuclides that have a split giant dipole resonance with two Lorentzian parameters sets, \texttt{RAINIER} takes the incoherent sum of the two GSFs.
Higher multipole order GSFs (M2, E3, M3, ...) are omitted since they have a very small influence on the cascade.
Where applicable, scissors or pygmy resonances have a standard Lorentzian form and low energy enhancement has the following soft-pole form:
\begin{equation}
 f_{enhance}(E_\gamma) = C_1 \textrm{exp}(-C_2 E_\gamma)
\label{eq:GSFEnhance}
\end{equation}
where $C_1$ and $C_2$ are experimentally fit parameters.

\subsection{\label{ssec:fluct}Width Fluctuations}

Equation \eqref{eq:avgWidth} is only a prescription for \textit{average} transition width.
An \textit{individual} width is defined by
\begin{equation}
\Gamma^{XL}_{I,F}(E,E_\gamma) = \frac{2 \pi}{\hbar} \lvert \braket{\Psi_F | H^{XL} | \Psi_I} \rvert^2,
\label{eq:transition}
\end{equation}
where $\braket{\Psi_F | H^{XL} | \Psi_I}$ is the matrix element of transition operator $H^{XL}$ connecting initial state wavefunction $\Psi_I$ and final state wavefunctions $\Psi_F$ which includes both residual nucleus and emitted photon.
In real nuclei, widths are hypothesized to independently fluctuate according to a width fluctuation distribution (WFD) commonly cast into a $\chi^2$ distribution:
\begin{equation}
P(x,\nu) = \nu/2 \cdot g(\nu/2)^{-1} \biggl(\frac{\nu x}{2} \biggr)^{\nu/2 - 1} e^{-\nu x / 2} 
\label{eq:chi2}
\end{equation}
where $x = \Gamma^{XL}(E,E_\gamma) / \bar{\Gamma}^{XL}(E,E_\gamma)$ is the ratio of a given width to the average, $g$ is the mathematical Gamma function, and $\nu$ is the number of degrees of freedom inherent to the system.
To simulate fluctuations, \texttt{RAINIER} sets each width equal to the calculated average from Equation \eqref{eq:avgWidth} multiplied by a random sample from the WFD in Equation \eqref{eq:chi2}.

By far the most widely used WFD is the Porter-Thomas Distribution (PTD) with $\nu = 1$ \cite{PhysRev.104.483}:
\begin{equation}
P(x, \nu = 1 ) = \frac{e^{-x/2}}{\sqrt{2\pi x}},
\end{equation}
which is the default WFD in \texttt{RAINIER}.
The PTD is equivalent to a Gaussian distribution squared.
This form has a physical explanation \cite{PhysRev.104.483}: the matrix element of Equation \ref{eq:transition} is equal to an integral over a multipole operator between two wavefunctions which are presumably unrelated to one another due to the complexity of strong interaction.
Therefore one may expect the matrix element probability distribution to be Gaussian with zero mean. 
However, recent results from Koehler \textit{et.\ al.\ }\cite{PhysRevLett.105.072502} show that the WFD may be more akin to a distribution with $\nu \approx 0.5$, suggesting that there may be more symmetry in the system.
\texttt{RAINIER} users have the option to set the parameter $\nu$ to a real-valued number greater than zero. 
This addition allows further tests of Random Matrix Theory to experimental situations where there is direct access to width fluctuations.

A given width can be many orders greater or lower than the average with non-trivial probability.
The chaos of these violent fluctuations is one of the primary motivations behind statistical decay programs like \texttt{RAINIER}: users can see the full effect of width fluctuations on experimental observables.
Fortunately, when there are many decay paths, these violent fluctuations tend to average out and give pseudo-stable values for observables such as low-lying level populations.
\texttt{RAINIER} has the potential to quantify the magnitude of width fluctuations effects by simulating many realizations of the randomly constructed level scheme. 

\subsection{\label{ssec:flow}Cascade Quantities and Program Flow}

After \texttt{RAINIER} determines an entrance state, it calculates the total transition width out of that state:
\begin{equation}
\Gamma_{I,tot} = \sum_{F,XL} \Gamma^{XL}_{I,F},
\label{eq:totWid}
\end{equation}
where the width sum includes possible transition types $XL$, to final states $F$, below the initial energy bin $I$.
Decay widths incorporate both $\gamma$-ray emission width, $\Gamma_\gamma^{XL}$, and electron internal conversion:
\begin{equation}
\Gamma^{XL}_{I,F} = \Gamma_\gamma^{XL} \cdot [1 + \alpha^{XL}(E_\gamma) ],
\label{eq:alpha}
\end{equation}
where $\alpha^{XL}$ are relativistic internal conversion coefficients from the evaluated $BrIcc$ tables \cite{KIBEDI2008202}.
\texttt{RAINIER} users can choose between the Frozen Orbitals, No-Hole, and all other approximations available in $BrIcc$.
\texttt{RAINIER} interpolates values from these tables, where the user can specify the range and number of interpolation points. 
Internal conversion coefficients are large in high $Z$ nuclei and in transitions near the K-edge, so more interpolation points are recommended in these areas.

Transition widths can be recast into other dimensionless forms such as branching ratios.
The branching ratio to a specific final state is a ratio of a particular decay channel width to the total width of the state:
\begin{equation}
\textrm{BR}_{I,F} = \sum_{XL} \frac{ \Gamma_{I,F}^{XL} }{\Gamma_{I,tot}}.
\label{eq:BR}
\end{equation}
\texttt{RAINIER} determines and records whether the chosen decay branch emits a $\gamma$-ray or an electron using their respective $BR$'s according to Equations \eqref{eq:alpha} and \eqref{eq:BR}.

During the decay calculations, \texttt{RAINIER} also computes multipole mixing,
\begin{equation}
\delta^2 = \abs{ \frac{ \mel{\Psi_F|}{E2}{|\Psi_I} } 
  { \mel{\Psi_F|}{M1}{|\Psi_I} } }^2
  = \frac{ \Gamma_{I,F}^{E2} }{ \Gamma_{I,F}^{M1} },
\end{equation}
where $\delta$ is limited to E2/M1 admixtures.
Furthermore, \texttt{RAINIER} computes lifetimes proportional to the reciprocal of total width,
\begin{equation}
\tau = \frac{\hbar} {\Gamma_{I,tot}}.
\label{eq:lifetime}
\end{equation}
Finally, \texttt{RAINIER} tracks cumulative event time where each decay step takes a random sample of time from an exponential distribution parameterized by the lifetime of the state:
\begin{equation}
P(t) = e^{-t / \tau} / \tau.
\label{eq:Ptau}
\end{equation}

\subsection{\label{ssec:codeStruct} Code Structure}

Transition widths have pseudorandom values that must remain consistent during repeated use within a realization.
Poorly treated width fluctuations are most noticeable when level spacings are large and exit channels are few. 
To handle fluctuations properly, \texttt{RAINIER} uses pseudorandom number generators (PRNGs) and initialization numbers known as \textit{seeds}.

\texttt{RAINIER} initializes the \textit{width}-governing PRNG with a seed based on the current level and realization number so that different levels and realizations have completely independent exit channels.
The \textit{event}-governing PRNG independently supplies random samples for initial state population, exit branch selection, decay time, and internal conversion calculations.
\texttt{RAINIER} initializes the event-governing PRNG with a seed based on the event and realization number.

When users operate the \textit{single state} population mode described in Section \ref{ssec:ipop}, \texttt{RAINIER} saves the primary decay widths. 
In other population modes, \texttt{RAINIER} recalculates all decay options upon entering a state.
There are often too many decay possibilities to save all transitions to memory. 
For instance, a charged particle reaction with a nucleus of $A \sim 150$ might have $10^4$ possible initial states, where each state has $10^7$ E1, M1, and E2 primary decay options, each comprising 8 bytes of memory for a total disk space of $\approx$$800$ G.
Moreover, simulations do not utilize all decay quantities in a reaction that has $10^{14}$ decay paths.
Even if \texttt{RAINIER} saved all decay quantities to memory, generating and looking up values in a large table would be overly time-consuming.

By default, \texttt{RAINIER} uses the Tausworthe PRNG \cite{L'Ecuyer:1996:MEC:228695.228713}, while the more widely-known Mersenne Twister (MT19937) \cite{Matsumoto:1998:MTE:272991.272995} is also available.
The PRNG state variables of MT19937 occupy 32 times more memory than the Tausworthe generator and thus take more initialization time for each entered state.
Statistical decay is a somewhat unique application of random numbers that requires frequent reinitialization of PRNGs and is rarely in danger of exceeding the $10^{26}$ period of the Tausworthe generator.


Most nuclear physics codes are written in \texttt{FORTRAN}, but \texttt{RAINIER} is written in the more modern \texttt{C++} programming language coupled to the \texttt{ROOT} framework \cite{Antcheva20092499} which is familiar to the experimental nuclear physics community.
Users can plot $\gamma$-ray spectra, known low-lying discrete level populations, feeding time distributions, and various other correlated observables without writing additional software.
The combination of quick plotting and readable source code makes \texttt{RAINIER} accessible to the average experimentalist.

\section{\label{sec:output}Output Examples}

\texttt{RAINIER} is versatile in its simulation input and output. 
This section provides an overview of \texttt{RAINIER}'s capabilities to simulate various types of reactions and extract experimentally observable quantities.

\subsection{\label{ssec:TSC}Two Step Cascade Spectra} 

\begin{figure}
\centerline{\includegraphics[width=0.75\linewidth]{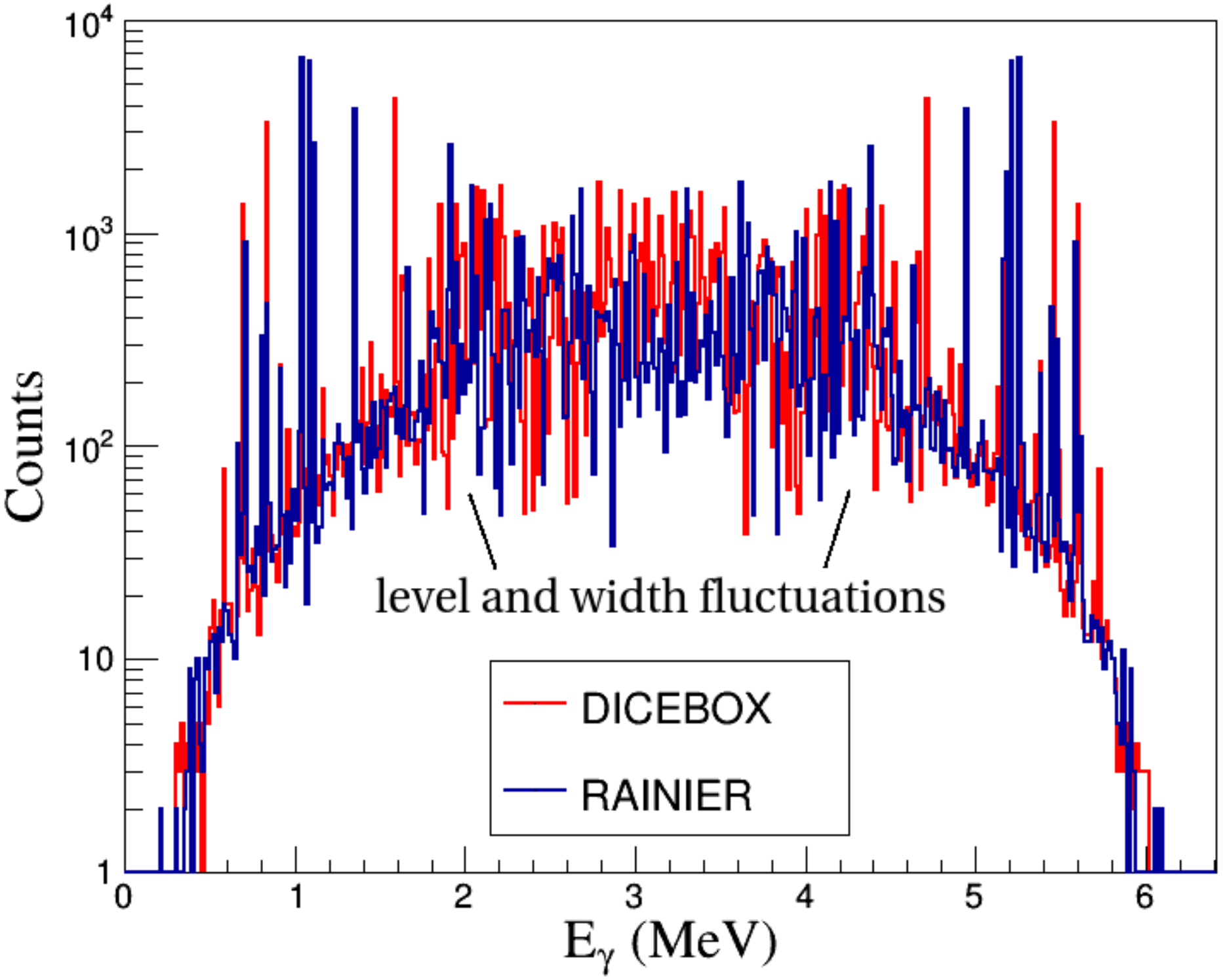}}
\caption{$^{143}$Nd(n,$\gamma$) to $3^-_1$ Two Step Cascade spectra benchmark of \texttt{RAINIER} to \texttt{DICEBOX}.
The overall energy dependence has excellent agreement between the codes. 
Intensity fluctuations differ as expected from dissimilar PRNGs.}
\label{fig:TSC}
\end{figure}

Spectra of two-step $\gamma$-ray cascades (TSC) following thermal neutron capture have been influential in determining NLD and GSF models for the past several decades \cite{PhysRevC.46.1276}.
The primary tool to deconvolve these spectra has been \texttt{DICEBOX} \cite{Bečvář1998434}.
In an effort to benchmark \texttt{RAINIER} to \texttt{DICEBOX}, Figure \ref{fig:TSC} shows TSC spectra from both codes for $^{143}$Nd(n,$\gamma$) with $^{144}$Nd input parameters similar to those of Reference \cite{PhysRevC.46.1276}:
\begin{itemize}
  \item 14 low-lying levels from Reference \cite{TULI1989607}
  \item Capture state: $3^-$ at $7.8174$ MeV
  \item Low-lying tagged final state: $3^-_1$ at $1.5109$ MeV 
  \item NLD model: BSFG with $a = 14.58$ MeV$^{-1}$ and $E_1 = 0.968$ MeV
  \item Spin cutoff model: Equation \eqref{eq:spinLowE} low-energy version 
  \item E1 GSF model: generalized Lorentzian of Equation \eqref{eq:GSFGen} with $S_{E1} = 317.0$ mb, $G_{E1} = 5.30$ MeV, and $E_{E1} = 15.05$ MeV
  \item M1 GSF model: standard Lorentzian of Equation \eqref{eq:GSFStd} with $S_{M1} = 0.37$ mb, $G_{M1} = 4.00$ MeV, and $E_{M1} = 7.82$ MeV
  \item E2 GSF model: single particle constant strength of $4 \times 10^{-11}$ MeV$^{-5}$
\end{itemize}
The internal conversion coefficient model was turned off ($\alpha = 0$) for consistency purposes; the number of conversion electrons is less than 2\% of events for this reaction.
The one realization shown has $10^6$ (n,$\gamma$) events.
For this simulation and those following, other defaults mentioned previously in this article are used including the parity equipartition model, Poissonian level spacing distribution, and $\nu = 1$ Porter-Thomas width fluctuations.

A more thorough benchmarking was performed using total widths, $\Gamma_{I,tot}$, near the neutron binding energy of 7.817 MeV.
The most significant difference that emerged between \texttt{DICEBOX} and \texttt{RAINIER} was the technique for random number generation from a Poisson distribution; the transition from a true Poissonian to a Gaussian approximation was inherent to the PRNG function call. 
The differing approximations resulted in a mere 0.2\% total width discrepancy in the most severe scenario.

\subsection{\label{ssec:drTSC}Direct Reaction Two Step Cascade} 

To further demonstrate the capabilities of \texttt{RAINIER}, consider recent results of Wiedeking \textit{et.\ al.\ }of $^{94}$Mo(d,p)$^{95}$Mo which provided confirmation of the low energy enhancement in the GSF \cite{PhysRevLett.108.162503}.
Their method uses a variant of the TSC method involving Direct Reactions (DRTSC), where particle energy provides the initial excitation energy of the residual nucleus and $\gamma$-ray transitions from low-lying levels specify the discrete states being fed.
The relative probabilities to decay from the set of particle-tagged energies to various low-lying levels of the same $J\Pi$ depends solely on the energy dependence of the GSF (see the original Letter \cite{PhysRevLett.108.162503} for more details).
\texttt{RAINIER} can reproduce this scenario using the \textit{spread of states} population mode.

\begin{figure}
\centerline{\includegraphics[width=\linewidth]{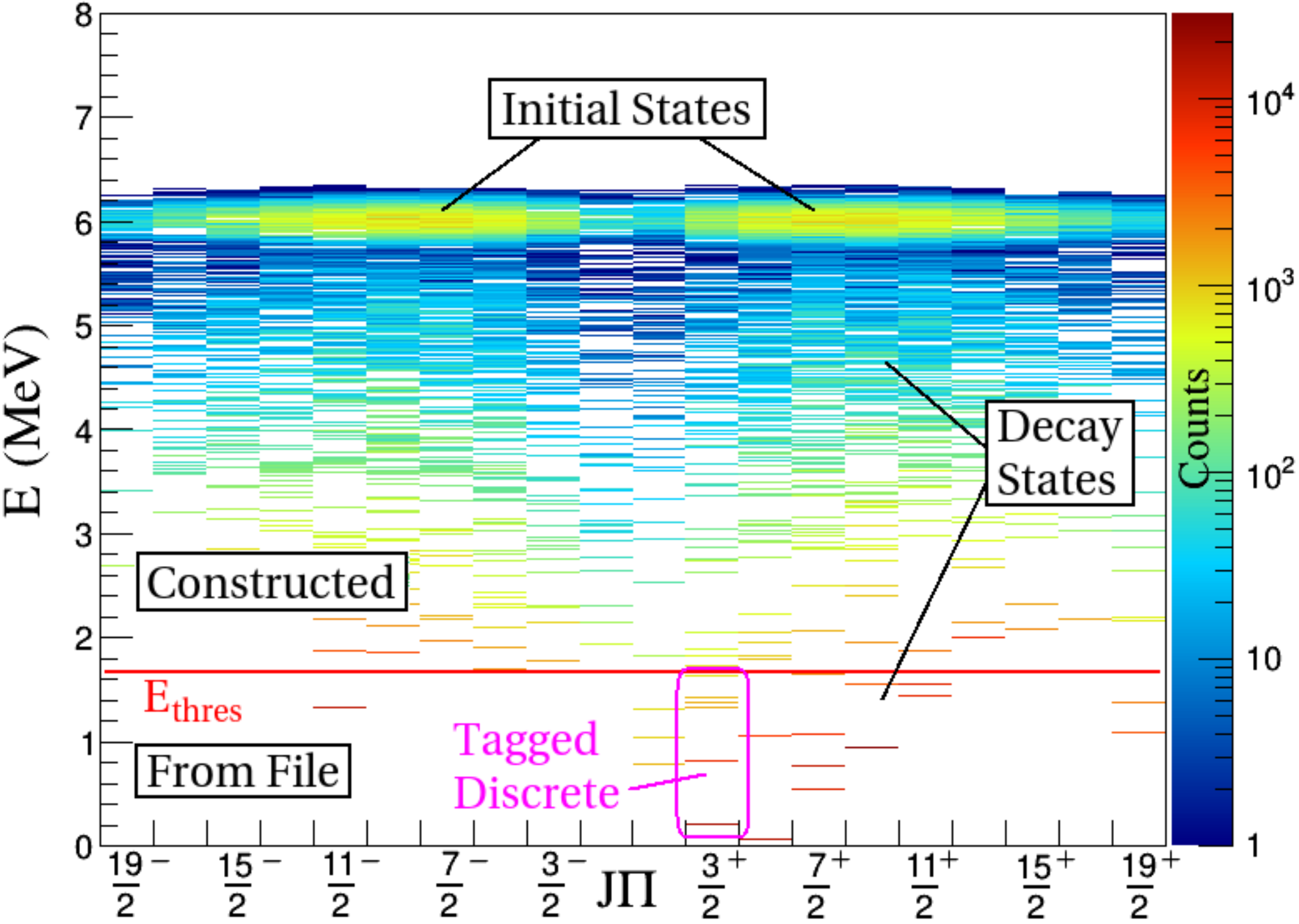}}
\caption{$^{95}$Mo excited to $\bar{E}_{I} = $ 6 MeV and subsequent decay to low-lying constructed states and states from the discrete level file.
}
\label{fig:depopulation}
\end{figure}

The following \texttt{RAINIER} simulation uses standard values from RIPL-3 for $^{95}$Mo NLD and GSF parameters.
The simulation inputs are the following:
\begin{itemize}
  \item NLD model: BSFG with $a = 9.78$ MeV$^{-1}$ and $E_1 = -0.42$ MeV
  \item Spin cutoff model: Equation \eqref{eq:spinLowE} low-energy version 
  \item E1 GSF model: generalized Lorentzian with $S_{E1} = 195.7$ mb, $G_{E1} = 5.488$ MeV, and $E_{E1} = 16.482$ MeV
  \item M1 GSF model: standard Lorentzian of Equation \eqref{eq:GSFStd}: $S_{M1} = 0.749$ mb, $G_{M1} = 4.00$ MeV, and $E_{M1} = 8.986$ MeV
  \item M1 GSF low energy enhancement model: soft pole of Equation \eqref{eq:GSFEnhance} with $C_1 = 7 \times 10^{-8} $ MeV$^{-3}$ and $C_2 = 2.0$ MeV$^{-1}$
  \item E2 GSF model: standard Lorentzian of Equation \eqref{eq:GSFStd} with $S_{E2} = 0.15$ mb, $G_{E2} = 4.97$ MeV, and $E_{E2} = 13.807$ MeV
  \item 24 low-lying levels with $J\Pi$ assignment corrections from the original DRTSC Letter \cite{PhysRevLett.108.162503}
  \item Five initial mean excitation energies: $\bar{E}_{I} = $ 3, 4, 5, 6, and 7 MeV with Gaussian resolution of 0.2 MeV
  \item Six $\gamma$-ray tagged low-lying states with $J\Pi = 3/2+$: $E =$ 0.204, 0.821, 1.370, 1.426, 1.620, and 1.660 MeV
  \item Internal conversion model: $BrIcc$ Frozen Orbital approximation \cite{KIBEDI2008202}
\end{itemize}
According to the original DRTSC Letter \cite{PhysRevLett.108.162503}, the $J_I$ distribution is insignificant provided there are a sufficient number of decays to the low-lying states of interest.
Figure \ref{fig:depopulation} shows initial state population and decay for $\bar{E}_{I} = 6 $ MeV and highlights the tagged low-lying discrete states.

\begin{figure}
\centerline{\includegraphics[width=0.85\linewidth]{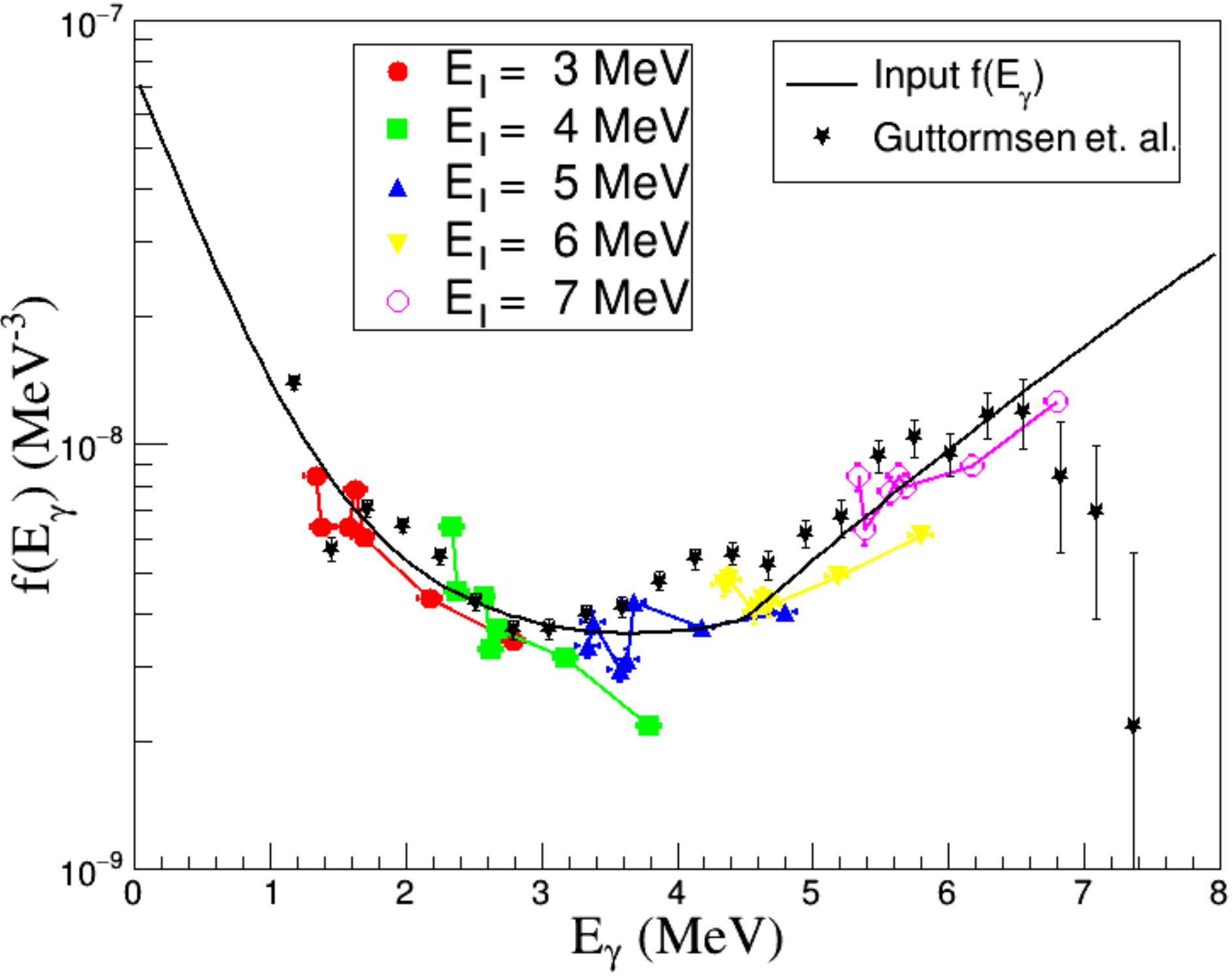}}
\caption{Simulated extractions of the energy dependence of the GSF using the DRTSC method of Wiedeking \textit{et.\ al.\ }\cite{PhysRevLett.108.162503} normalized to experimental data of Guttormsen \textit{et.\ al.\ }\cite{PhysRevC.71.044307}.
}
\label{fig:extract}
\end{figure}

Figure \ref{fig:extract} shows the results of the DRTSC extraction method of the GSF energy dependence. 
The DRTSC method cannot determine absolute magnitude of the GSF, so each set of primary decays must be normalized to experimental data.
Figure \ref{fig:extract} includes experimental data and the simulation input of total GSF:
\begin{equation}
f(E_\gamma) = f_{E1}(E_\gamma, E = 5 \textrm{ MeV}) + f_{M1}(E_\gamma).
\end{equation}

Level spacing and transition width fluctuations will distort the reliability of the DRTSC extraction procedure at sufficiently low level densities.
A large number of entrance states averages out width fluctuations, whereas a small number of entrance states would have more scatter in the extracted data points.
Figure \ref{fig:extract} suggests that 0.2 MeV wide excitation bins in $^{95}$Mo are suitable for extracting the energy dependence of the GSF without severe fluctuations.
\subsection{\label{ssec:oslo}Oslo Method} 

The Oslo method \cite{GUTTORMSEN1987518} analyzes a large range of experimental nuclear excitations, from the maximum kinematically allowable excited state all the way down to the first excited state.
Data is typically collected from a charged particle reaction where silicon detectors determine $E_{I}$ from the ejectile kinetic energy and an array of scintillators measure the $\gamma$-ray spectra.
Emission of $\gamma$-rays depends on the behavior of the NLD and GSF at all energies as well as on the initial and underlying $J$ distributions.
The following example uses \texttt{RAINIER} to simulate $E_{I}$ vs.\ $E_\gamma$ output of 16 MeV $^{56}$Fe(p,p') and performs the Oslo method to extract NLD and GSF.

\begin{figure*}
\centerline{\includegraphics[width=0.75\linewidth]{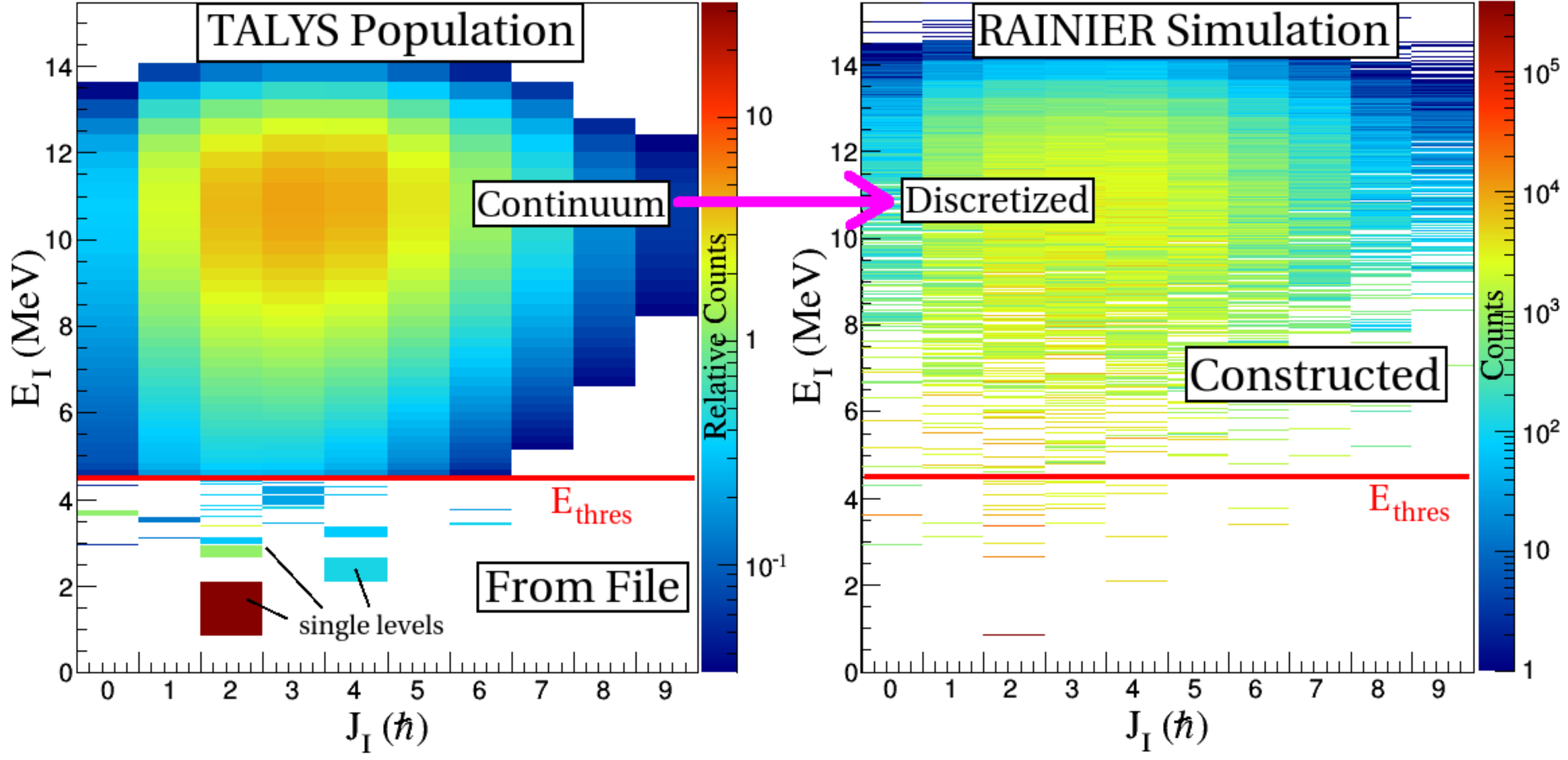}}
\caption{Population of $^{56}$Fe initial states from 16 MeV (p,p'). 
\texttt{RAINIER} randomly samples the continuous distribution of \texttt{TALYS} output above $E_{thres}$ and selects the nearest discrete level of the constructed level scheme.}
\label{fig:PopTALYS}
\end{figure*}

Since $^{55}$Fe is unstable, there is no $^{55}$Fe(n,$\gamma$)$^{56}$Fe data for the $^{56}$Fe level density and total radiative width at the neutron binding energy.
Thus the following \texttt{RAINIER} simulation uses systematic formulae compiled in RIPL-3 for $^{56}$Fe NLD and GSF.
The simulation inputs are the following:
\begin{itemize}
  \item NLD model: BSFG with $a = 5.854$ MeV$^{-1}$ and $E_1 = 1.0715$ MeV
  \item Spin cutoff model: Equation \eqref{eq:spinEmpirical} empirical fit with $Pa' = 2.905$ MeV
  \item E1 GSF model: KMF Lorentzian of Equation \eqref{eq:GSFKMF} with $S_{E1} = 91.519$ mb, $G_{E1} = 6.976$ MeV, and $E_{E1} = 18.687$ MeV
  \item M1 GSF model: standard Lorentzian of Equation \eqref{eq:GSFStd} with $S_{M1} = 1.101$ mb, $G_{M1} = 4.00$ MeV, and $E_{M1} = 10.717$ MeV
  \item M1 GSF model: low energy enhancement soft pole of Equation \ref{eq:GSFEnhance} with $C_1 = 4 \times 10^{-7} $ MeV$^{-3}$ and $C_2 = 2.5$ MeV$^{-1}$
  \item E2 GSF model: standard Lorentzian of Equation \eqref{eq:GSFStd} with $S_{E2} = 0.075$ mb, $G_{E2} = 5.438$ MeV, and $E_{E2} = 16.467$ MeV
  \item 31 low-lying levels from RIPL-3
  \item Internal conversion model: $BrIcc$ Frozen Orbital approximation \cite{KIBEDI2008202}
\end{itemize}
The \textit{full reaction} population mode of \texttt{RAINIER} is used in this simulation with the initial $E J \Pi$ distribution originating from \texttt{TALYS} output.
The \texttt{TALYS} keyword ``outpopulation'' invokes an output section titled \textit{Population of $^{56}$Fe Before Decay} for the $E J$ histogram.
Equal populations of positive and negative parity are assumed above $E_{thres}$.
In practical terms, \texttt{RAINIER} performs the final stage processing of the reaction using statistical $\gamma$-ray decay methods.
Figure \ref{fig:PopTALYS} shows \texttt{TALYS} continuum populations and a conversion by \texttt{RAINIER} to discrete populations. 

\begin{figure*}
\centerline{\includegraphics[width=0.75\linewidth]{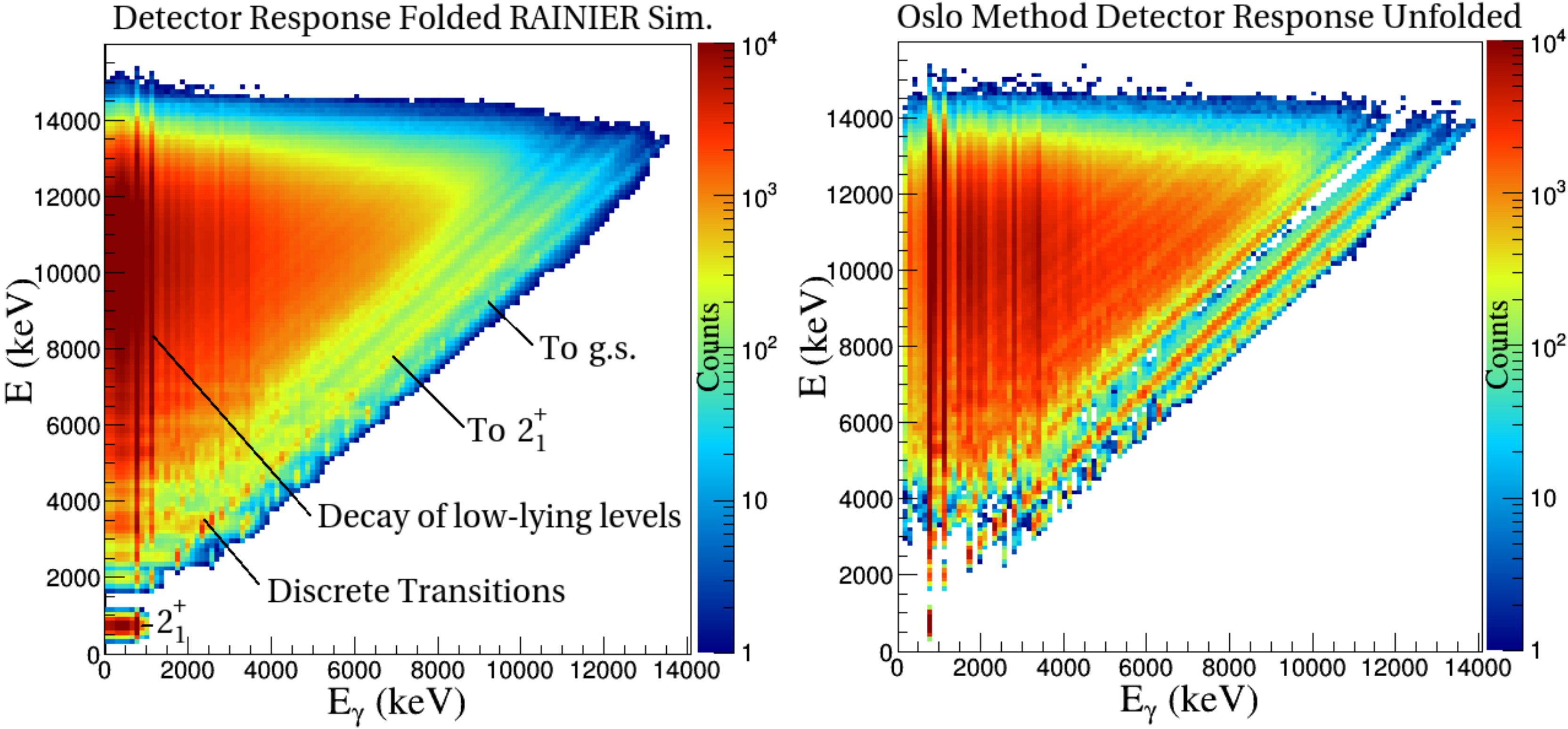}}
\caption{Left: \texttt{RAINIER} simulated depopulation of 16 MeV $^{56}$Fe(p,p') via $\gamma$-ray cascades with CACTUS array detector response and 0.2 MeV excitation energy resolution.
Right: $\gamma$-ray spectrum after detector response unfolding.
}
\label{fig:Oslo}
\end{figure*}

Figure \ref{fig:Oslo} shows results of the $\gamma$-ray cascade depopulation of the $E J \Pi$ bins as a function of excitation energy.
To make this simulation as near to real experimental conditions as possible, an excitation energy resolution of 0.2 MeV has been included as well as the response function of the University of Oslo's CACTUS NaI detector array. 
Many familiar features appear such as primary decay to low-lying levels from all excitation energies and secondary low-lying discrete transitions.
Figure \ref{fig:Oslo} also shows the results of the Oslo method detector response unfolding technique \cite{GUTTORMSEN1996371}. 

\begin{figure*}
\centerline{\includegraphics[width=0.75\linewidth]{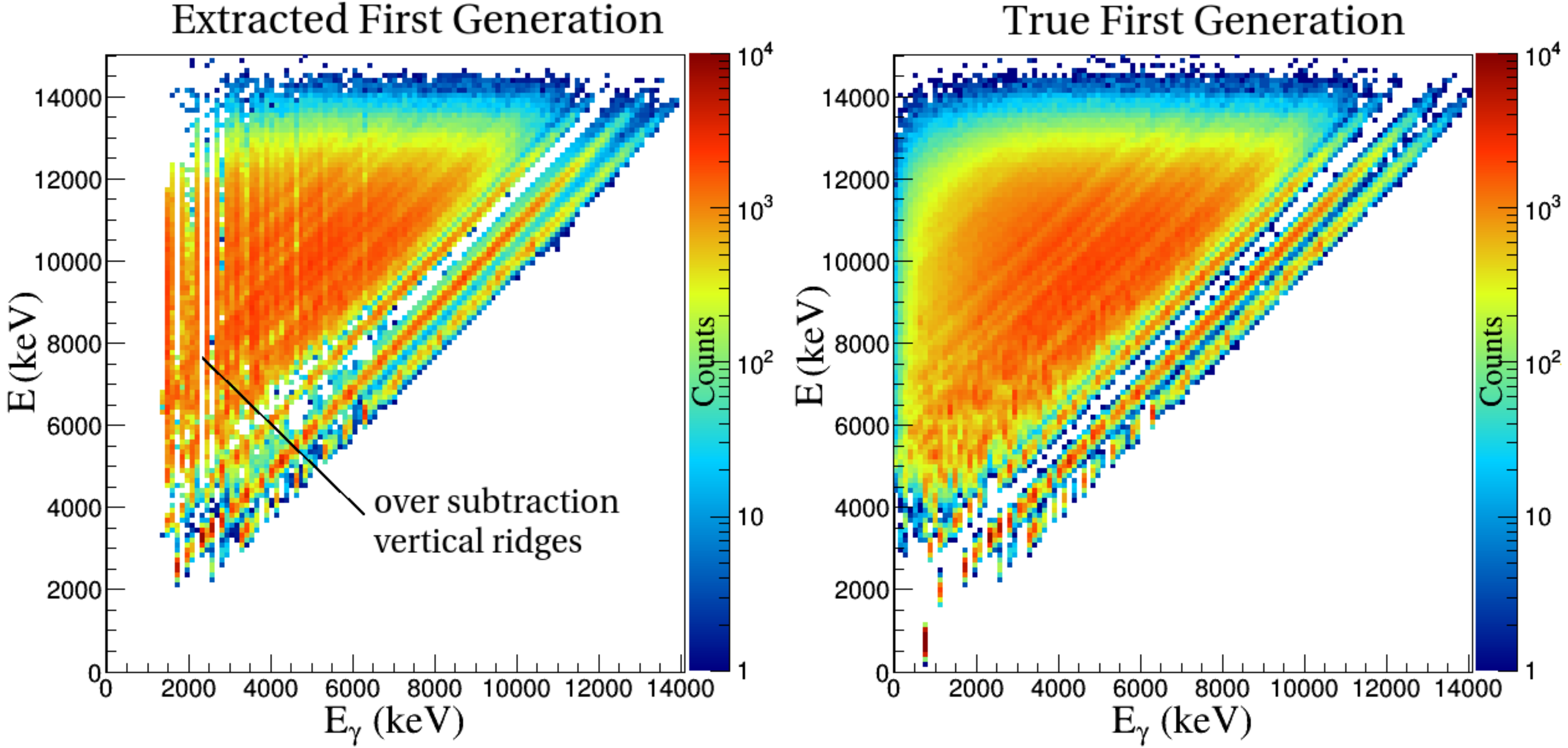}}
\caption{Left: $\gamma$-ray spectrum after application of 0.2 MeV excitation energy resolution, detector response folding, unfolding, and extraction of the first generation spectrum.
Right: true \texttt{RAINIER} first generation $\gamma$-ray spectrum.
The vertical lines in the unfolded first generation extracted spectrum are the result of over-subtraction of low-lying transitions that are strongly populated from high excitation energies.}
\label{fig:fg1Gen}
\end{figure*}

Figure \ref{fig:fg1Gen} shows results of the Oslo method first generation extraction procedure \cite{GUTTORMSEN1987518} applied to the unfolded spectrum of Figure \ref{fig:Oslo}.
Figure \ref{fig:fg1Gen} also shows the true first generation spectrum directly from the \texttt{RAINIER} simulation devoid of $\gamma$-ray detector response unfolding and first generation extraction operations (i.e.\ no smoothing from detector resolution and no artifacts from the extraction procedure). 
The overall comparison between the unfolded first generation extracted spectrum and the true first generation is satisfactory. 
This agreement confirms that the unfolding and extraction procedures adequately recover the primary $\gamma$-ray spectrum in spite of multiple $\gamma$-ray emission, level spacing and transition width fluctuations, and detector uncertainty.
At low $E_\gamma$, there are some vertical lines in the unfolded first generation extracted spectrum that are not present in the true first generation spectrum.
These vertical ridges are the result of an over-subtraction of $\gamma$-ray transitions out of a state that is strongly populated in the decay cascades of high excitation energies but that is only moderately populated via direct excitation.
The intensity and placement of these vertical ridges will vary from realization to realization.
This fluctuation interferes with the extracted NLD and GSF in a real measurement where there is only one true distribution of levels and widths.

\begin{figure*}
\centerline{\includegraphics[width=0.75\linewidth]{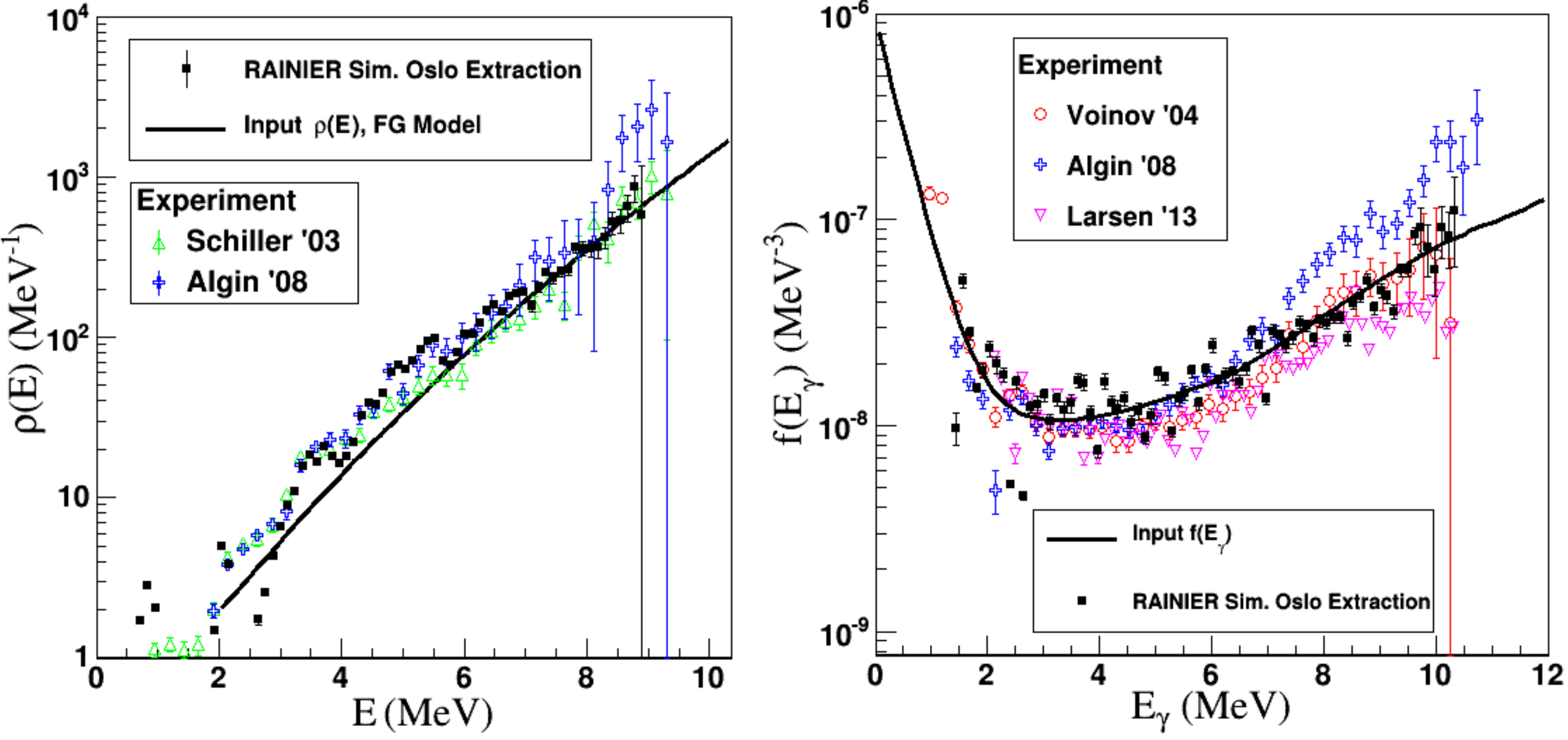}}
\caption{NLD and GSF extracted using the full Oslo method on \texttt{RAINIER} simulated data. Shown for reference are the simulation input of NLD and GSF as well as a comparison of experimental Oslo method extractions for $^{56}$Fe. }
\label{fig:ExtractNLDGSF}
\end{figure*}

Figure \ref{fig:ExtractNLDGSF} shows results of the final stage of the Oslo method to extract NLD and GSF from the unfolded first generation extracted spectrum of Figure \ref{fig:fg1Gen}.
Fluctuations in NLD and GSF extracted from simulated data are comparable to fluctuations in NLD and GSF extracted from other experiments \cite{PhysRevLett.93.142504,PhysRevLett.111.242504,PhysRevC.78.054321,PhysRevC.68.054326}.
Note that the Oslo method is sufficiently sensitive at low $E_\gamma$ to extract the enhancement in GSF as first seen by Schiller \textit{et.\ al.\ }\cite{PhysRevC.68.054326}. 
The low energy enhancement feature disappears from the Oslo method extraction results if the \texttt{RAINIER} simulation excludes the enhancement in the GSF input.


\subsection{\label{ssec:pop}Angular Momentum Distributions}

It is important to emphasize the difference between the underlying and initial angular momentum distributions.
The underlying angular momentum distribution is universal to all reactions with a specified target nucleus and is a fundamental nuclear property governed by the spin cutoff parameter in Equations \eqref{eq:spDist}-\eqref{eq:spinEmpirical}.
The initial angular momentum ($J_I$) distribution depends on the projectile particle type and incident energy and can be experimentally manipulated during data analysis as explained in Section \ref{ssec:ipop}.
The $J_I$ distribution is difficult to determine theoretically due to the complexity of competing direct, compound, and pre-equilibrium nuclear formation mechanisms.
While the $J_I$ distribution is often unknown, it can be experimentally deduced if needed for additional calculations and simulations. 
Comparisons of experimental and simulated low-lying level populations can help estimate the \textit{mean} of the initial angular momentum distribution, $\bar{J}_I$.

Consider again the 16 MeV $^{56}$Fe(p,p') reaction where an outgoing proton is detected at $\approx 6$ MeV and intensities of detected signature $\gamma$-rays determine low-lying level populations.
\texttt{RAINIER} can reproduce these low-lying populations to help deduce $\bar{J}_I$.
The following simulation uses the same $^{56}$Fe nuclear input parameters as Section \ref{ssec:oslo} with the exception that the initial state population method has been changed to the \textit{spread of states} mode with the following properties:
\begin{itemize}
\item One initial mean excitation energy: $\bar{E}_{I} = $ 10 MeV with Gaussian resolution of 0.2 MeV
  \item Initial angular momentum distribution: Poissonian with mean $\bar{J}_I = 3.5, 4.5 $ $\hbar$
\item Uniform initial parity distribution
\end{itemize}
The simulation output will elucidate which low-lying level populations are most sensitive to $\bar{J}_I$.
An experimenter can then measure those level populations to determine $\bar{J}_I$.

\begin{figure}
\centerline{\includegraphics[width=0.9\linewidth]{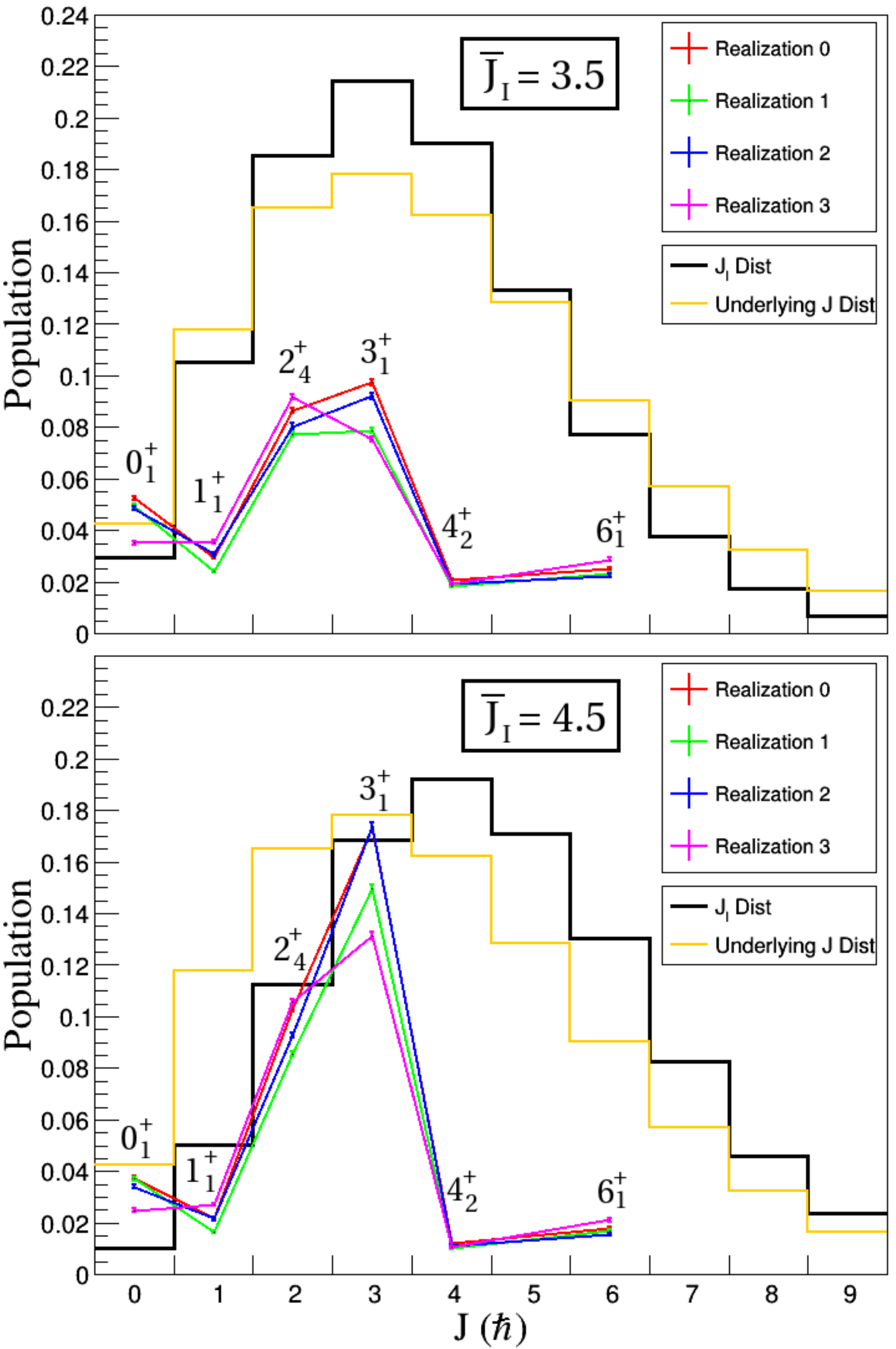}}
\caption{
Simulated low-lying level population in $^{56}$Fe with $\bar{E}_{I} = 10$ MeV. 
Top: $\bar{J}_I = 3.5$ $\hbar$. 
Bottom: $\bar{J}_I = 4.5$ $\hbar$.
The underlying $J$ distribution is inherent to the nucleus, while the $J_I$ distribution is reaction dependent.
Low-lying level populations are instrumental to determine $\bar{J}_I$.
}
\label{fig:SpPop}
\end{figure}

Figure \ref{fig:SpPop} shows simulated low-lying populations of the $0^+_1$, $1^+_1$, $2^+_4$, $3^+_1$, $4^+_2$, and $6^+_1$ in $^{56}$Fe for four realizations and the two different $\bar{J}_I$.
This simulation requires multiple realizations because level spacing and width fluctuations introduce variations in low-lying populations on the order of 20\% as shown in the figure.
The most sensitive levels to $\bar{J}_I$ are the $2^+_4$ and $3^+_1$.
The populations of the $2^+_4$ and $3^+_1$ are roughly equal when $\bar{J}_I = 3.5$ $\hbar$, but the population of the $3^+_1$ is always larger than the $2^+_4$ when $\bar{J}_I = 4.5$ $\hbar$ regardless of realization.
Hence for this reaction and low-lying level populations, an experimenter can determine $\bar{J}_I$ with accuracy of at least 1 $\hbar$ without worrying about the influence level spacing and width fluctuations.

\subsection{\label{ssec:feed}Quasi-Continuum Lifetimes and Feeding Time Distributions} 
Theoretically, lifetimes at high excitation energies depend solely on NLD, the GSF, and fluctuations in level spacings and transition widths as related by equations \eqref{eq:lvlEJP}, \eqref{eq:lvlSpac}, \eqref{eq:avgWidth}, \eqref{eq:chi2}, and \eqref{eq:lifetime}.
In principle, Doppler shift measurements can indicate the amount of time elapsed between excited state formation and low-lying level population yielding some information about the magnitude of quasi-continuum lifetimes (QC$\tau$), which subsequently yields information about the magnitude of NLD and GSF at high excitation energies.
No previous publication known to these authors and collaborators attempts to simulate feeding time distributions or report experimental measurements of QC$\tau$.

The following \texttt{RAINIER} simulation uses the same $^{56}$Fe nuclear input parameters as Section \ref{ssec:oslo} with the exception that the initial state population method has been changed to the \textit{spread of states} mode with the following properties:
\begin{itemize}
  \item Five initial mean excitation energies: $\bar{E}_{I} = $ 6, 7, 8, 9, and 10 MeV with Gaussian resolution of 0.2 MeV
  \item Initial angular momentum distribution: Poissonian with mean $\bar{J}_I = 3.5 $ $\hbar$
  \item $4 \times 10^6$ events per mean excitation energy
\end{itemize}
Width fluctuations were absent in the population of levels, but they were present in the $\gamma$-ray decay.

\begin{figure}
\centerline{\includegraphics[width=0.9\linewidth]{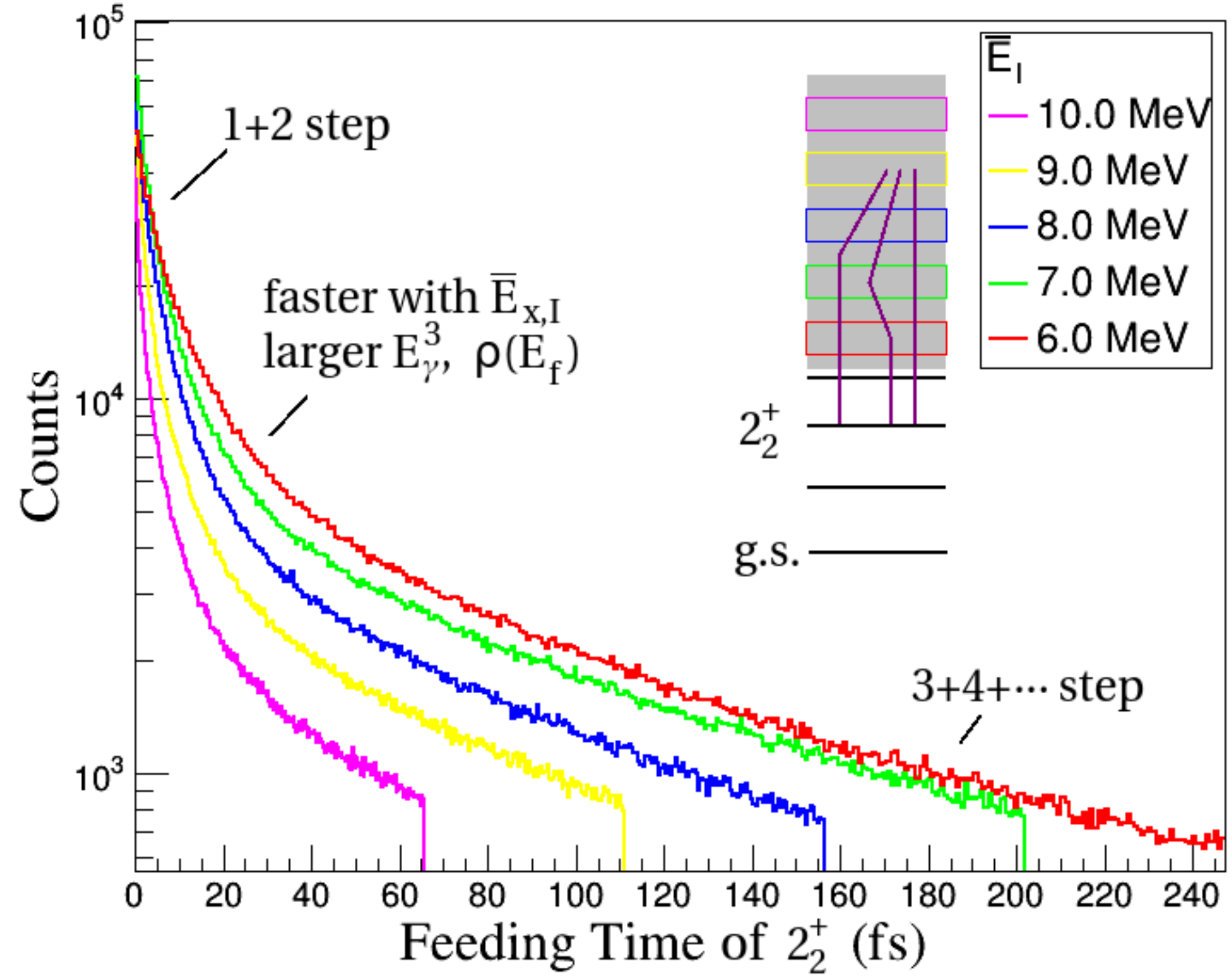}}
\caption{Simulated feeding time distributions to the $2^+_2$ in $^{56}$Fe 
from various initial excitation energies, $\bar{E}_{I}$.
The illustration depicts the 9 MeV initial excitation energy range decaying directly to the $2^+_2$ or via a series of intermediate steps.
}
\label{fig:feed}
\end{figure}

Figure \ref{fig:feed} shows simulation results of feeding time distributions to the $2^+_2$ state in $^{56}$Fe for the various initial excitation ranges. 
As indicated by a steeper slope in counts with time, higher energy initial excitations decay faster because more exit states are available and the large factor of $E_\gamma^{2 L + 1}$ significantly increases transition widths. 
As indicated by the fewer number of total counts, population intensity of the $2^+_2$ state decreases with excitation energy because there are more ways to bypass the state in the decay chain.

The feeding time distribution of Figure \ref{fig:feed} is only the first step toward simulation of experimental observables involving Doppler shift.
Traditionally, measurements using the Doppler Shift Attenuation Method (DSAM) \cite{Alexander1978} rely on knowledge of nuclear recoil trajectories, where excited nuclei recoil through target material and slow down via interactions with electrons and other nuclei.
Usually a Monte Carlo simulation \cite{WINTER1983537} is necessary to handle the severe and erratic changes of the recoil velocity vector as a function of time.
These simulations incorporate the simple exponential decay curve of Equation \eqref{eq:Ptau}, where lifetime of a discrete low-lying state is the sole parameter for the time distribution.
However, the feeding time distributions of Figure \ref{fig:feed} are far more complex than a one parameter fit because the feeding process involves multi-step cascades with all intermediate levels.
Quantities such as \textit{average feeding time} are not very meaningful to the DSAM technique because there is a point at which the nucleus is fully stopped and any additional time does not influence Doppler shift.
Therefore, one must incorporate the full feeding time distribution into the list of recoil trajectories to compare with experimental Doppler shifts.
This integration of feeding time with recoil trajectory is the subject of a future article including a comparison of the simulations to experimental output.


\section{\label{sec:concl}Conclusion}

\texttt{RAINIER}, a new program introduced here, adds cascade fluctuations to the simulation of $\gamma$-ray decay from a wide range of initial nuclear states. 
Previous reaction code packages such as \texttt{TALYS} and \texttt{EMPIRE} populate and decay a range of states, but neglect fluctuations.
Conversely, the program \texttt{DICEBOX} includes fluctuations, but populates only only two initial states.
New experimental techniques populating a wide range of states require simulation incorporating level spacing and transition width fluctuations to understand the volatility of their analysis methods.
For some nuclei, recent results from the Oslo method show an enhancement in the GSF at low $E_\gamma$ \cite{PhysRevC.68.054326} and a scissors resonance when N$>$Z \cite{PhysRevLett.41.1532}.
As the field of nuclear physics moves to measurements farther from the valley of stability with the Facility for Rare Isotope Beams \cite{FRIBtechRep} and the Gamma-Ray Energy Tracking In-beam Nuclear Array \cite{Paschalis201344}, the familiar models of NLD and GSF may further transform in unpredictable ways.
NLDs and GSFs of nuclei far from stability remain important inputs for many applications since these quantities govern the balance between particle and $\gamma$-ray emission.
For instance, neutron capture cross sections determine reaction rates in stellar nucleosynthesis \cite{PhysRevC.82.014318} and nuclear fission reactors \cite{refId0}.

This paper has shown several \texttt{RAINIER} simulation examples including TSC spectra, the direct reaction TSC method, the Oslo Method, low-lying $J$ populations, and finally feeding time distributions.
Application of \texttt{RAINIER} will allow new experimental methods of NLD and GSF extraction to confirm their findings with simulation and to test the resilience of their techniques to level spacing and width fluctuations.
Furthermore, \texttt{RAINIER} is prepared to test new models of NLD and GSF as the Facility for Rare Isotope Beams and the Gamma-Ray Energy Tracking Array come online to probe nuclei far from the valley of stability.

The \texttt{RAINIER} source code is hosted on \url{https://github.com/LEKirsch/RAINIER} for public access and it is written in \texttt{C++} with prolific explanatory comments for user-friendly readability. 
\ref{sec:sampleRun} gives a quick overview of the code layout including an example of how to run it.


\section{\label{sec:ackn}Acknowledgements}

This work was performed with the support of the DOE NNSA Stewardship Science Graduate Fellowship under cooperative agreement number DE-NA0002135.

It is a pleasure to thank M. Krti\v{c}ka for stimulating discussion and maintaining \texttt{DICEBOX}, A. Hurst for providing an excellent introduction to statistical decay, M. Guttormsen and A. C. Larson for experimental enthusiasm, A. Ureche for provoking questions, and A. Lewis for reviewing the manuscript and code.

\appendix
\section{\label{sec:sampleRun}Example of Running \texttt{RAINIER}}

This section provides a quick, brief overview for starting \texttt{RAINIER}.
If you do not have \texttt{ROOT} \cite{Antcheva20092499} installed, try the following installation command for Linux systems 
\begin{verbatim}
sudo apt install root-system-bin
\end{verbatim}
or go to the \texttt{ROOT} download website: \url{https://root.cern.ch/downloading-root}.

\begin{figure}
\centerline{\includegraphics[width=\linewidth]{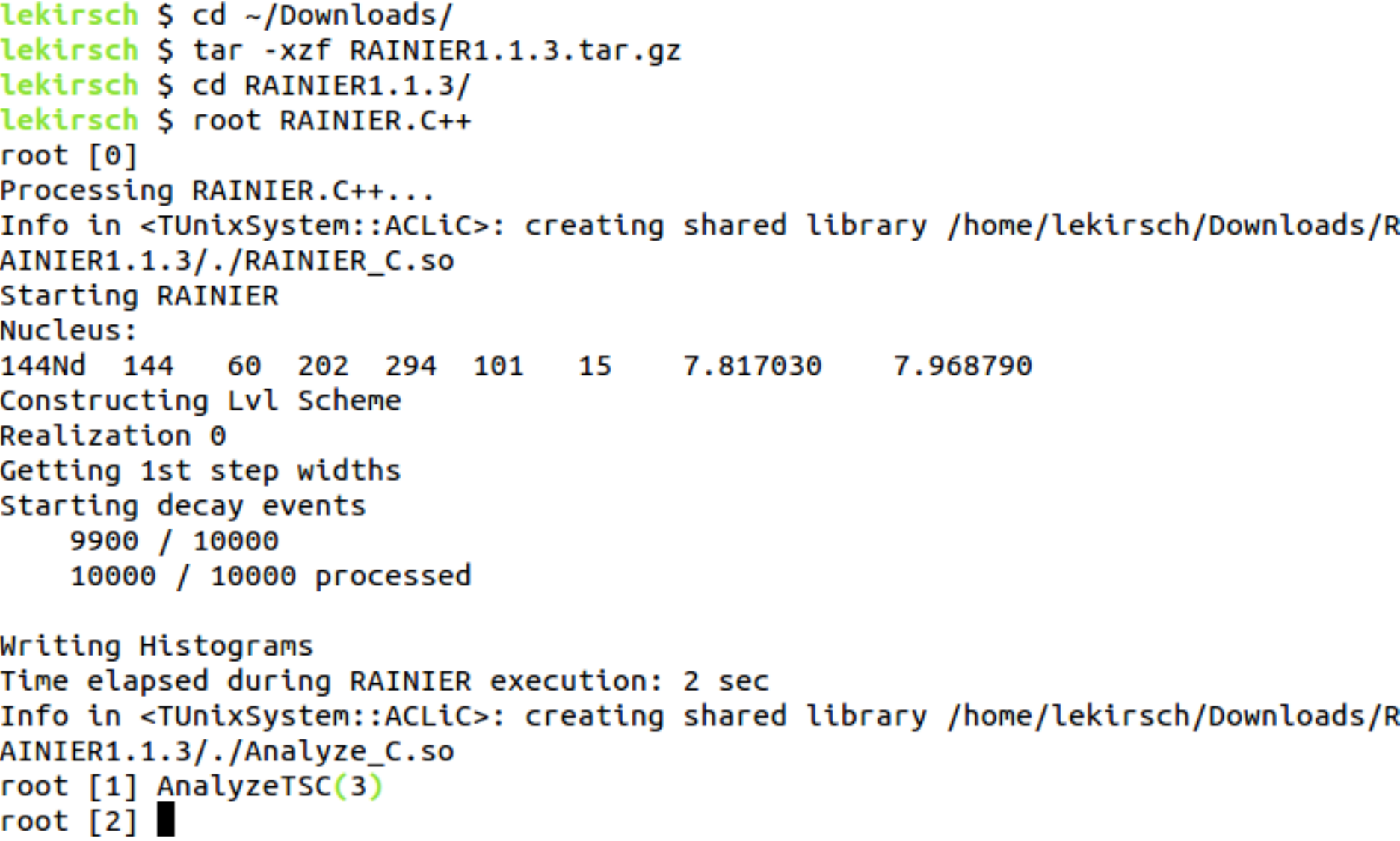}}
\caption{After downloading the distribution tarball, generating your first spectra takes just five easy commands.
This sequence will produce the \texttt{RAINIER} portion of the simulation in Figure \ref{fig:TSC}.}
\label{fig:Run}
\end{figure}

Next, download the latest \texttt{RAINIER} distribution package from from \url{https://github.com/LEKirsch/RAINIER}.
Unzip the tarball as shown in Figure \ref{fig:Run}.
The $BrIcc$ slave program \texttt{briccs} \cite{KIBEDI2008202} is included in this package and new versions will be included in future distributions.
Execute \texttt{RAINIER} within the \texttt{RAINIER}\textit{version}\texttt{/} directory with the following bash command:
\begin{verbatim}
root RAINIER.C++
\end{verbatim}
After the program ends, the function \texttt{AnalyzeTSC(int nDisEx)} will plot the Two Step Cascade spectrum to the discrete level specified by \texttt{nDisEx}.

\begin{figure}
\centerline{\includegraphics[width=\linewidth]{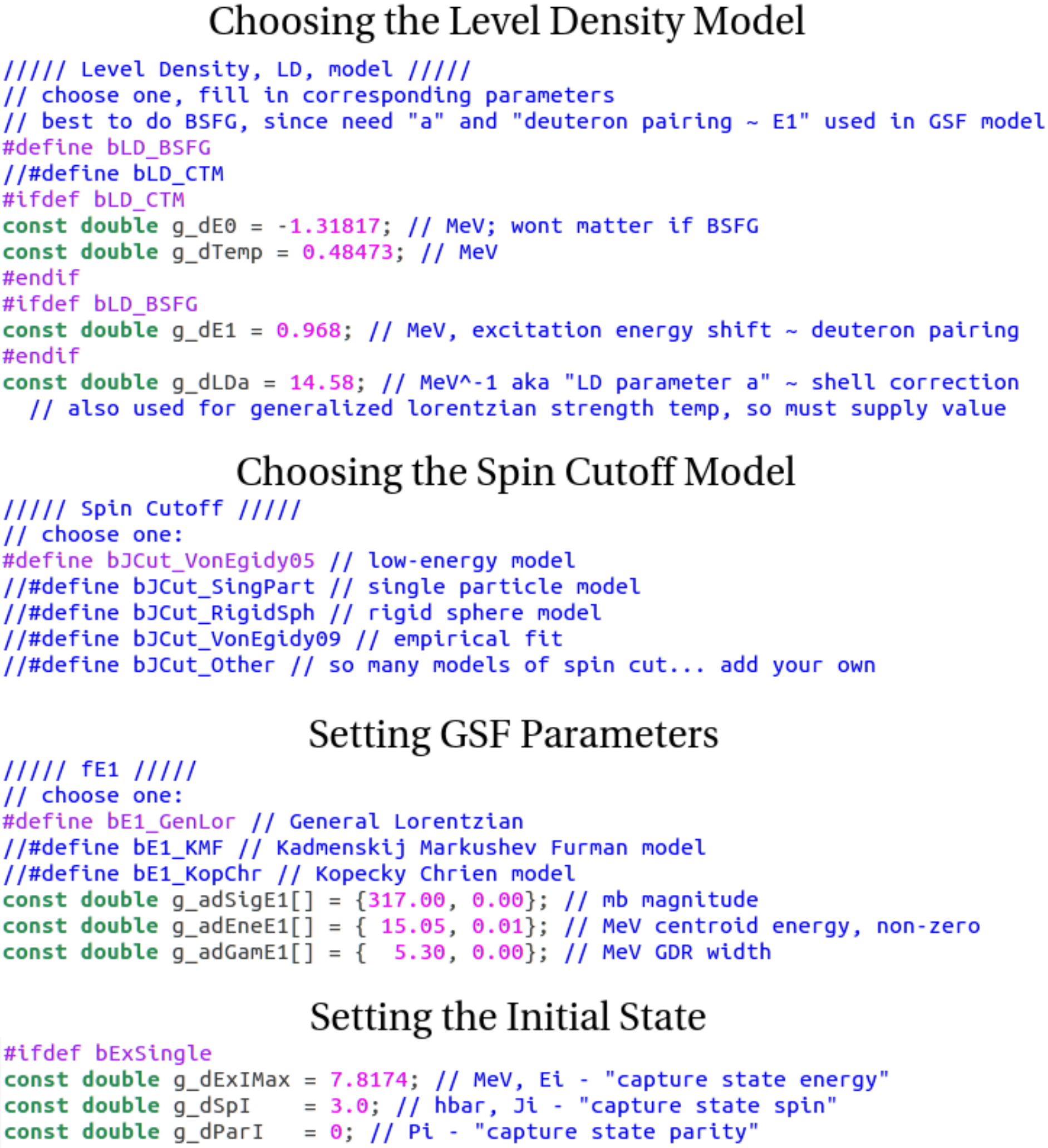}}
\caption{Several sections of the TSC simulation input file for Figure \ref{fig:TSC}.
Numerous comments provide a guide to understanding the input parameters and programmatic flow.
}
\label{fig:Input}
\end{figure}

To modify the input to meet the specifications of your experiment, open the \texttt{RAINIER.C} file and edit the basic parameters shown in Figure \ref{fig:Input}.
NLD, GSF, spin cutoff, and initial state population parameters change most regularly from experiment to experiment.
Other essentials like $A$, $Z$, and the discrete level file must also be updated.

Further manipulations of \texttt{RAINIER} input include changing the number of degrees of freedom, $\nu$, in the WFD, changing the level spacing distribution to incorporate Wigner fluctuations, and changing the internal conversion model lookup table of $BrIcc$.
Adding additional physics models, initial state distributions, and level scheme details is within the capabilities of the average programmer.

\bibliography{pub}

\end{document}